\documentclass[letterpaper,twocolumn,superscriptaddress,floatfix]{revtex4}

\usepackage[latin1]{inputenc}
\usepackage{bm}
\usepackage{multirow,amssymb,amsbsy,amsmath}
\usepackage{graphicx}
\usepackage{verbatim}
\makeatletter
\usepackage{pifont}
\makeatother

\begin{document}

\title{Quantum storage of three-dimensional orbital-angular-momentum entanglement in a crystal}

\author{Zong-Quan Zhou}
\author{Yi-Lin Hua}
\author{Xiao Liu}
\author{Geng Chen}
\author{Jin-Shi Xu}
\author{Yong-Jian Han}
\author{Chuan-Feng Li$\footnote{email:cfli@ustc.edu.cn}$}
\author{Guang-Can Guo}
\affiliation{Key Laboratory of Quantum Information, University of Science and Technology of China,
CAS, Hefei, 230026, China}
\affiliation{Synergetic Innovation Center of Quantum Information and Quantum Physics, University of Science and Technology of China, Hefei, 230026, P.R. China}
\date{\today}
\begin{abstract}
{Here we present the quantum storage of three-dimensional orbital-angular-momentum photonic entanglement in a rare-earth-ion-doped crystal. The properties of the entanglement and the storage process are confirmed by the violation of the Bell-type inequality generalized to three dimensions after storage ($S=2.152\pm0.033$). The fidelity of the memory process is $0.993\pm0.002$, as determined through complete quantum process tomography in three dimensions. An assessment of the visibility of the stored weak coherent pulses in higher-dimensional spaces, demonstrates that the memory is highly reliable for 51 spatial modes. These results pave the way towards the construction of high-dimensional and multiplexed quantum repeaters based on solid-state devices. The multimode capacity of rare-earth-based optical processor goes beyond the temporal and the spectral degree of freedom, which might provide a useful tool for photonic information processing.}
\end{abstract}
\pacs{03.67.Bg, 03.67.Hk, 42.50.Md, 32.80.Qk} 
\maketitle

Photons are the natural carriers of information in quantum networks. However, because photon loss scales exponentially with the channel length, quantum communication protocols that utilize the direct transmission of photons are limited to distances of approximately hundreds of kilometers \cite{repeaterreview}. To overcome such limitations, the concept of a quantum repeater has been proposed that utilizes entanglement swapping and quantum memory to efficiently distribute long-distance entanglement \cite{BDCZ}. The storage of entangled photon pairs is a fundamental requirement of such a quantum repeater, and many experimental efforts have been devoted to the realization of such light-matter interfaces in recent years \cite{timebin1,timebin2,polar1,polar2,oam}. The light-matter interface can be further utilized to perform advanced processing tasks \cite{njp,Reim12,lgi} which may find applications in both quantum and classical information processing.

Currently, the memory lifetime and the achievable data rate represent two major challenges towards the practical realization of quantum-repeater-based quantum networks \cite{repeaterreview}. It is possible to extend the former to a satisfied minute scale as already demonstrated for storage of classical light in a solid-state quantum memory \cite{life}. Two specific methods have been proposed to enhance the data rate; one is the utilization of high-dimensional encodings \cite{cryp,capacity}, and the other is the use of multimode memories \cite{mm,repeaterreview,mmt,modes3}. The orbital-angular-momentum (OAM) of a photon has proven to be an outstanding degree of freedom (DOF) for carrying high-dimensional entanglement \cite{cryp,capacity,qbc,coin} and for spatial multimode operations \cite{oammm}. Significant progress has been made through the storage of OAM states of classical light \cite{oamc1} and single photons \cite{oamp1,oamp2,oamp3}. Recently, the quantum storage of two-dimensional OAM entanglement has been reported in a quantum memory based on cold atomic gases \cite{oam}. For quantum memory to become practical, the experimental implementation based on solid-state systems is appealing because of the reduced complexity of setup and high robustness. Here, we experimentally demonstrate that the quantum memory based on a rare-earth-ion (RE) -doped crystal, which can faithfully store three-dimensional OAM entanglement and provide large spatial-multimode capacity, is a promising way towards the construction of a practical quantum repeater. The faithful storage of spatial information may significantly improve the information processing capacity of RE-based optical processors.

\begin{figure*}[tb]
\centering
\includegraphics[width=0.65\textwidth]{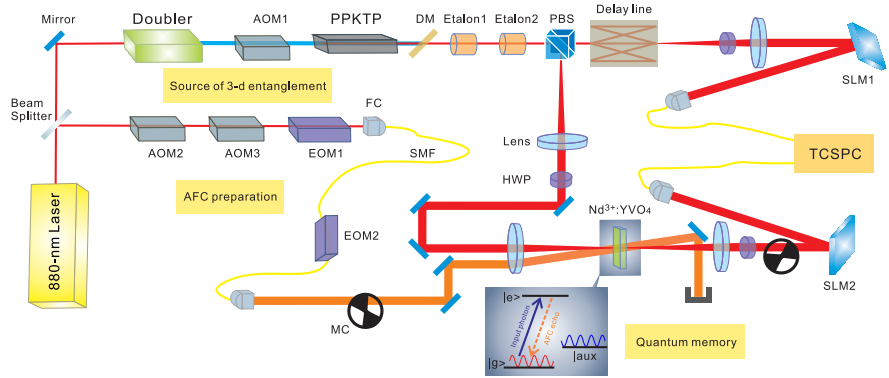}
\caption{\label{Fig:1} (color online) Schematic diagram of the experimental setup. The photon pairs are generated in the 20-mm PPKTP crystal via SPDC. The pump source is obtained by sending the master laser to a frequency doubler. AOM1 and AOM2 are used to ensure the frequency matching between the AFC and the center frequency of the photon pairs. The blue pump light is then removed by the dichroic mirror (DM). The photon pairs are spectrally filtered by the two etalons. The two photons in pair are separated by the polarization beam splitter (PBS). The signal photon of each pair is directed to the memory hardware; the idler photon is forced to propagate through a 16.4-m delay line and are then analyzed with a spatial light modulator (SLM) and a single-mode fiber (SMF). AOM3 and the two electro-optic modulators (EOMs) together generate the pump light for AFC preparation. The polarization of the signal photons is controlled by the half-wave plates (HWPs). The AFC preparation light and the signal photons overlap at the sample position, with a small angle (20 mrad) between them. The Nd$^{3+}$:YVO$_4$ crystal is placed in a cryostat (Oxford Instruments, SpectromagPT) with a temperature of 1.5 K and a magnetic field of 0.3 T. After a programmable delay, the OAM of the retrieved signal photons is analyzed using the SLM2 and a SMF. The single-photon signals are analyzed using time-correlated single-photon counting system (TCSPC). The phase-locked mechanical choppers (MC) are used to protect the single-photon detector (SPD) during the preparation procedure. The inset shows the energy structure of Nd$^{3+}$ ions. The AFC is prepared in the ground state $|$g$\rangle$ by frequency-selective optical pumping of atoms from $|$g$\rangle$ to an auxiliary Zeeman state $|$aux$\rangle$. An input photon resonant with the $|$g$\rangle\longrightarrow|
$e$\rangle$ transition will be stored as a collective excitation of the AFC. The AFC works as a grating in the frequency domain and diffracts the photon into a temporal-delayed echo emission. Further details are provided in Ref. \cite{SI}.}
\end{figure*}

\begin{figure}[tb]
\centering
\includegraphics[width=0.4\textwidth]{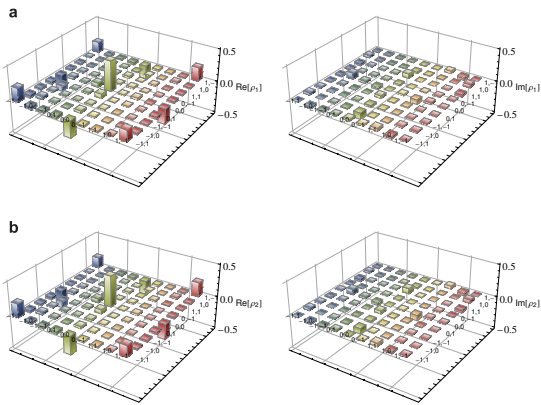}
\caption{\label{Fig:2} (color online) Graphical representation of the reconstructed density matrix of the two photon states before (a) and after (b) the storage process. The density matrix of the two qutrits is reconstructed from a set of 81 measurements represented by the operators $u_i\otimes u_j$ (with $i,j=1,2,...,9$) and $u_k=|\psi_k\rangle\langle\psi_k|$). The kets $|\psi_k\rangle$ for both the idler photons and the signal photons are selected from the following set: \{$|-1\rangle, |0\rangle, |1\rangle, (|0\rangle+|-1\rangle)/\sqrt{2}, (|0\rangle+|1\rangle)/\sqrt{2}, (|0\rangle+i|-1\rangle)/\sqrt{2}, (|0\rangle-i|1\rangle)/\sqrt{2}, (|-1\rangle+|1\rangle)/\sqrt{2}, (|-1\rangle+i|1\rangle)/\sqrt{2}$\}. Detailed descriptions of the tomographic measurements are presented in Ref. \cite{SI}.}
\end{figure}

\begin{table*}
\renewcommand{\arraystretch}{1.2}
\textbf{Table 1. Entanglement measures.}\\
\begin{tabular}{|c|c|c|c|c|}
\hline
  & Negativity & Fidelity to the MES & Input/output fidelity & Measured S\\\hline
$\rho_1$ & $0.643\pm0.005$ & $0.730\pm0.006$ & / &$2.150\pm0.030$\\\hline
$\rho_2$ & $0.656\pm0.010$ & $0.742\pm0.011$ & $0.991\pm0.003$& $2.152\pm0.033$\\\hline
\end{tabular}
\caption{Entanglement measures. The negativity is calculated as $(||\rho^{T_B}||_1-1)/2$ where $||\rho^{T_B}||_1$ is the trace norm of $\rho^{T_B}$ and $\rho^{T_B}$ is the partial transpose of a state $\rho$. The fidelity to the maximally entangled state (MES) is calculated as $[Tr(\sqrt{\sqrt{\rho}\rho_0\sqrt{\rho}})]^2$ where $\rho_0=|\psi_0\rangle \langle\psi_0|$ and $|\psi_0\rangle=(|-1\rangle|1\rangle-|0\rangle|0\rangle+|1\rangle|-1\rangle)/\sqrt{3}$.  The fidelity between the input and output states is calculated as $[Tr(\sqrt{\sqrt{\rho_2}\rho_1\sqrt{\rho_2}})]^2$}\label{table_1}
\end{table*}

First, we demonstrate the quantum storage of three-dimensional OAM entanglement in a spatial-multimode solid-state quantum memory. A schematic diagram of our experimental setup is presented in Fig. 1. Our photon pairs are produced via a type-II spontaneous parametric down-conversion (SPDC) process in a PPKTP crystal. The OAM states of the generated photon pairs will be entangled because of angular momentum conservation \cite{oamspdc2}. The photon pair generated in this manner will have the form
\begin{equation}
|\Psi\rangle=c_{-1}|-1\rangle|1\rangle+c_{0}|0\rangle|0\rangle+c_1|1\rangle|-1\rangle+...,
\label{s}
\end{equation}
where $|c_i|^2$ ($i$=-1, 0, 1) is the probability of creating a photon pair with related states and $|l\rangle$ represents the OAM eigenstate as defined by the Languerre-Gauss mode LG$_{0,l}$ \cite{OAM1}. Photons with considerably higher $l$ can also be created in this process. However, we restrict the state to three dimensions because these three components have the most significant contributions to the produced state \cite{SI}.

The initial bandwidth of the photons is approximately 200 GHz, as determined by the phase-matching requirements. The photons are further filtered by two etalons to achieve a bandwidth of 700 MHz. Note that the etalons are plane-parallel cavities in which all transverse modes are degenerate \cite{etalon}; therefore the spectral filtering techniques presented here are also applicable for the entanglement of higher dimensions. The quality of the entanglement for the directly generated photon pairs is low. Three procedures are employed to concentrate the qutrit-qutrit entanglement, namely, adjusting the lens spacings \cite{oamspdc2}, inserting the etalons and tuning the temperature of the PPKTP crystal \cite{SI}. The quantum state of the photon pair is then characterized via quantum state tomography \cite{qst,oamatom}. A graphical representation of the reconstructed density matrix $\rho_1$ of the photon source is presented in Fig. 2a.

The memory is based on the atomic frequency comb (AFC) protocol, which requires a tailored absorption profile with a series of periodic absorption peaks separated by $\Delta$. An incident photon will be stored as a collective excitation of the AFC and released after a time delay of $1/\Delta$ \cite{AFC09,timebin1,timebin2,polar2}. The memory hardware is a Nd$^{3+}$:YVO$_4$ crystal (5-ppm doping level, 3-mm thickness). The spectroscopic properties of the sample are available in \cite{lgi}. An AFC bandwidth of approximately 1 GHz is achieved in the experiment through the modulation of the light using one double-pass acousto-optic modulator (AOM3), one free-space phase electro-optic modulator (EOM) and one fiber-coupled phase EOM. A storage efficiency of approximately 20\% with a storage time of 40 ns is achieved for the filtered photons from SPDC. Details of AFC preparation and an example of coincidence histogram for AFC storage of OAM entanglement are given in the Supplemental Material (SM) \cite{SI}.

The reconstructed density matrix of the readout state $\rho_2$ is presented in Fig. 2b and is nearly identical to that presented in Fig. 2a. We adopt two entanglement measures to quantify the 3D entanglement, namely, the negativity and the fidelity to the maximally entangled state (MES). The results are presented in Table 1. The fidelity of the states before and after storage is $0.991\pm0.003$, which demonstrats that this quantum memory is highly reliable for storing the 3D entanglement. To further demonstrate that the observed correlation cannot be explained by local realistic models, we use a generalized type of Bell inequality for qutrits \cite{belld}. The Bell tests lie at the heart of quantum physics and serve as a standard method for the identification of quantum nonlocality. Using the reconstructed density matrix, we searched for the specific measurement settings required for violation of the inequality. The measured values of $S$ before and after the storage process are presented in Table 1. Violation by more than 4 standard deviations is achieved after the storage process. The violation of the Bell inequality in $3 \times 3$ dimensions directly indicates the genuine 3D entanglement of the photon source and the reliability of the quantum memory.

\begin{figure}[tb]
\centering
\includegraphics[width=0.5\textwidth]{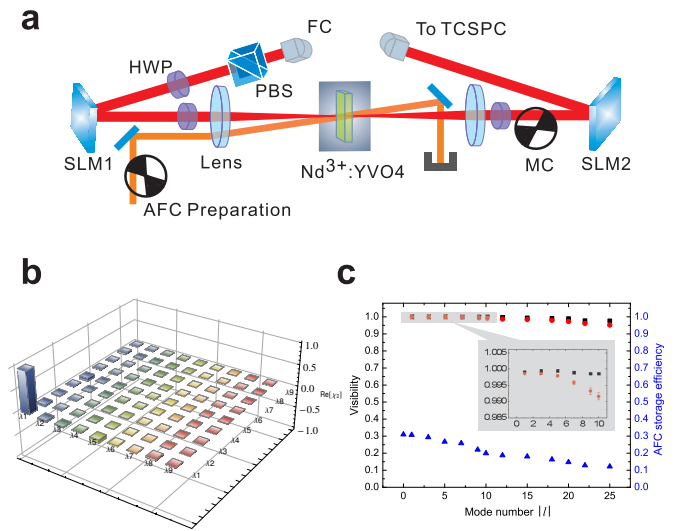}
\caption{\label{Fig:3} (color online) a. The setup for exploration of the multimode capacity in the spatial domain of the memory. The photon source is weak coherent pulse in Gaussian mode. The pulse has a temporal width of 20 ns and contains an average of 0.5 photons per pulse. Superposition states of OAM are generated though SLM1 and analyzed by SLM2. b. Graphical representation of the real part of the reconstructed process matrix $\chi_2$ in three dimensions. $\lambda_{1}$ is the identity operator, and the detailed definitions of the basis operators $\lambda_{1}\sim\lambda_{9}$ are provided in Ref. \cite{SI}. c. The memory performance for quantum superposition states $|\psi_+ (l)\rangle$. The black squares and red dots represent the visibility of the states before and after storage, respectively. The blue triangles represent the AFC storage efficiency for different values of $|l|$.}
\end{figure}

Now we turn to explore the high dimensionality and spatial-multimode capacity of the memory. A schematic illustration of the setup is presented in Fig. 3a.
Weak coherent pulses in the Gaussian mode are sent to SLM1. Single photons are converted into superposition states in high-dimensional spaces. The pulses that are incident on the sample contain an average of 0.5 photons per pulse. After storage in the memory, the retrieved photon states are analyzed using SLM2 and SMF. The two SLMs are set in the 4f configuration. To completely characterize the memory performance in three dimensions, we performed the quantum process tomography for qutrit operations \cite{qpt,SI}. We first measure the entire setup without the quantum memory; the reconstructed process matrix $\chi_1$ demonstrates a fidelity of $0.970\pm0.001$ with respect to an ideal quantum process $\chi_0$. Then, we measure the entire setup including the memory; the real part of the reconstructed process matrix $\chi_2$ is presented in Fig. 3b. For the ideal process, this matrix should be identical to $\chi_1$, meaning that the memory does not modify the input states but only delays it. The measured fidelity between $\chi_1$ and $\chi_2$ is $0.993\pm0.002$. These results indicate that although the setup for OAM analysis is not ideal, the memory performance itself is very close to ideal.

To assess the memory performance for states living in higher-dimensional spaces, the superposition states $|\psi_+ (l)\rangle = 1/\sqrt{2}(| l \rangle+ | -l \rangle)$ are prepared and stored in the memory. The readout photons are detected with $|\psi_+ (l)\rangle$ and $|\psi_- (l)\rangle=1/\sqrt{2}(| l \rangle - | -l \rangle)$. Examples of SLM pattern that generate these superposition states are given in the SM \cite{SI}. The storage performance can be assessed based on the visibility $V$ of the readout photons. $V$ is defined as $(P_+-P_-)/(P_++P_-)$, where $P_+$ and $P_-$ are the probabilities of detecting readout photons with states of $|\psi_+ (l)\rangle$ and $|\psi_- (l)\rangle$, respectively. In an ideal storage process, the state after storage would be exactly identical to $|\psi_+ (l)\rangle$ with $V=1$. The visibility of the readout photons is indicated with red dots, and results nearly identical to those for the input photons are obtained. The value of $V$ after storage is $0.952\pm0.008$ for $|l|=25$, demonstrating the high reliability of the memory for 51 spatial modes in total. The current pump power was optimized for storage efficiency of the $l=\pm1$ and $l=0$ modes, representing conditions similar to those for the storage of  three-dimensional entanglement. Note that the storage efficiency for higher $|l|$ can be significantly improved by simply increasing the pump power. For a quantum memory that is actually operating for 51 OAM modes, the storage efficiency should be balanced for various $|l|$ values with a single pump power. This can be achieved by saturating the pump power for lower $|l|$ at the expense of slightly lower efficiencies for all the modes. An example of such a configuration is presented in Fig. S10 in the SM \cite{SI}. Note that the balanced efficiency for various OAM modes is easily achieved in our solid-state quantum memory; however, it still represents a considerable challenge for quantum memory based on cold atomic gases \cite{oamp3}. A balanced and ideal storage efficiency for all modes can be expected with an ideal spatially uniform AFC. It can be prepared by utilizing a pump light with a super-Gaussian beam profile, which can be obtained by sending a Gaussian beam into an apodizing filter.

The large multimode capacity of the quantum memory should enable the construction of a quantum repeater with a higher achievable data rate and reduced memory time requirements \cite{mm,repeaterreview,mmt,modes}. On the other hand, if the quantum memory is used as a portable quantum hard drive with an extremely long lifetime \cite{6hours}, the storage capacity would be especially important. The AFC protocol has been proposed as a temporal multimode memory \cite{AFC09} and has been shown to possess a large multimode capacity in both the temporal \cite{modes} and spectral \cite{modes3} domain. As opposed to other memory protocols, the number of simultaneously storable modes in the AFC memory does not depend on the medium absorption depth \cite{Nunn08}. Considering the practical implementations, the temporal multimode capacity is determined by the ratio of the total AFC bandwidth to the bandwidth of each absorption peak and the spectral multimode capacity is determined by the ratio of the total AFC bandwidth to the bandwidth of each spectral mode. It is obvious that the temporal multimode capacity can be several times larger than the spectral multimode capacity, and that both of them are limited by the total AFC bandwidth. We note that the capacity of spatial multiplexing through the OAM DOF shows no dependence on AFC bandwidth; it is only determined by the spatial profile of the pump beam, which has no fundamental limitations provided that there are enough absorption centers and the medium is large enough. The RE-based memory is especially suitable for spatial multiplexing because it has a huge number of stable and active centers \cite{Ahlefeldt13} as well as unlimited transverse size. For practical implementations, one may expect more than 10$^6$ qubit modes \cite{qmp} in total by introducing the OAM modes with radial number and by utilizing the OAM mode sorters \cite{sorter,sorter2} to cooperate with the temporal and spatial multiplexing.

We have shown the storage of 3D OAM entanglement based on a two-level AFC in a RE-doped solid. There are two approaches to making the current memory useful in a quantum repeater. The first is DLCZ-like protocol, which requires on-demand quantum memory \cite{mm}. Then, the optical excitation should be further mapped into a spin state to achieve on-demand recall \cite{AFC09}. However, because another ground state is required to implement such three-level AFC protocol, one will need to move to Pr$^{3+}$- \cite{Gundogan15} or Eu$^{3+}$ \cite{spinecho}-doped solids with limited memory bandwidth of tens of MHz. To achieve a balanced and high efficiency of spin transfer for all OAM modes, the control pulse will be required to have increased diameter hence a significantly higher power. Better filtering techniques would be required to further suppress the scattering noise from the control beam. The isotope-enriched $^{143}$Nd$^{3+}$- or $^{145}$Nd$^{3+}$ -doped solids also can be considered as candidate systems with potentially larger storage bandwidth \cite{isotope}. Further spectroscopic measurements are required to identify a suitable ``$\Lambda$" system in such material. The other approach is based on spectral (spatial) multiplexing which requires fixed-time quantum memory \cite{modes3,fiber} and feed-forward control of readout. It is important that the memory should have high efficiency (90\%) with a storage time of hundreds of microseconds. The efficiency of the two-level AFC memory can approach unity by embedding the medium into a impedance-matched cavity \cite{Sabooni13}. The storage time is limited by the optical coherence time, which can be several milliseconds for RE-doped solids \cite{Sun02}. Therefore, this approach can be feasibly implemented through the combination of these techniques .

RE-doped solids are very promising candidates to permit implementation of both quantum \cite{Longdell04} and classical \cite{Harris98} information processing. The configurable beam splitter as well as the arbitrary temporal and spectral manipulations have been demonstrated with two-level AFC photonic processor in Ref. \cite{njp}; networks of such devices would allow scalable photonic information processing \cite{Reim12}. Encoding information in the spatial domain could significantly improve the information-processing capacity of such a photonic processor. Furthermore, a spatial beam splitter could be constructed based on SLM and OAM mode sorters so that spectral and/or temporal information would be combined into a selected spatial mode at will. As the AFC protocol is suitable for storage of intense light pulses, the storage of images may prove useful for classical image processing and image-correlation applications \cite{Gilboa04}.

The high fidelity of the storage process achieved here can be predominantly attributed to the fact that the ions contributing to the AFC storage are all spatially localized and minimal distortion of the spatial phase profile is introduced. Quantum memories based on solid-state systems have many practical advantages for future applications, including simple implementation, ease of manufacturing and integration, and unlimited medium size. Future efforts to combine high efficiency \cite{Sabooni13}, long storage time \cite{life,6hours,spinecho}, and multimode capacity in a single memory should lead to fascinating applications in both quantum and classical information processing.

\emph{Note added.}-Storage of high-dimensional OAM entanglement was also recently achieved in a quantum memory based on cold atomic gases \cite{ref}.

{\bf  Acknowledgments}
We thank Dong-Sheng Ding for helpful discussions. This work was supported by the National Natural Science Foundation of China (Nos. 61327901, 61490711, 11274289, 11325419, 11105135, 11474267, 61308010, 11274297 and 61322506), the National Basic Research Program of China (No. 2011CB921200), the Strategic Priority Research Program (B) of the Chinese Academy of Sciences (Grant No. XDB01030300), the Fundamental Research Funds for the Central Universities (Nos. wk2030380004 and WK2470000011).

\end{document}